%% file: main.tex
\documentclass[sigconf,prologue,dvipsnames]{acmart}

\usepackage[T1]{fontenc}
\usepackage{enumitem}
\usepackage{afterpage}
\usepackage{listings}
\usepackage{xcolor}
\usepackage{graphicx}
\usepackage{float}
\usepackage[font=small]{caption}
\usepackage[font=footnotesize]{subcaption}
\usepackage{placeins}
\usepackage{url}
\usepackage{epstopdf}
\usepackage{booktabs}
\usepackage{footnote}
\makesavenoteenv{table}
\makesavenoteenv{tabular}
\usepackage{flushend}

\usepackage[acronyms,nonumberlist,nopostdot,nomain,nogroupskip]{glossaries}

\usepackage{tabularx}

\usepackage{balance}

\definecolor{codegray}{rgb}{0.25,0.25,0.25}
\definecolor{codepurple}{rgb}{0.58,0,0.82}
\lstdefinestyle{mystyle}{
  commentstyle=\color{PineGreen},
  keywordstyle=\color{MidnightBlue},
  numberstyle=\tiny\color{codegray},
  stringstyle=\color{codepurple},
  basicstyle=\ttfamily\footnotesize,
  breakatwhitespace=true,         
  breaklines=true,                 
  captionpos=b,
  frame=tb,
  keepspaces=true,                 
  numbers=left,                    
  numbersep=5pt,                  
  showspaces=false,                
  showstringspaces=false,
  showtabs=false,                  
  tabsize=2,
  xleftmargin=10pt,
  belowskip=-10pt,
  float=htbp,  
}

\lstdefinelanguage{mybash}{%
  language     = bash,
  morekeywords = {docker,python3},
}

\input{acronyms.tex}

\input{myTikz.tex}

\begin{document}


\flushbottom
\setlength{\parskip}{0ex plus0.2ex}
\addtolength{\skip\footins}{-0.2pc plus 40pt}

\title{ns-O-RAN: Simulating O-RAN 5G Systems in ns-3}

\author{Andrea Lacava}
\affiliation{%
   \institution{Northeastern University}
   \city{Boston}
   \state{Massachusetts}
   \country{USA}}
\additionalaffiliation{%
   \institution{University of Rome, Rome, Italy}}
\email{lacava.a@northeastern.edu}

\author{Matteo Bordin}
\affiliation{%
   \institution{Northeastern University}
   \city{Boston}
   \state{Massachusetts}
   \country{USA}}
\email{bordin.m@northeastern.edu}

\author{Michele Polese}
\affiliation{%
   \institution{Northeastern University}
   \city{Boston}
   \state{Massachusetts}
   \country{USA}}
\email{m.polese@northeastern.edu}

\author{Rajarajan Sivaraj}
\affiliation{%
   \institution{Mavenir, Inc.}
   \city{Richardson}
   \state{Texas}
   \country{USA}}
\email{rajarajan.sivaraj@mavenir.com}

\author{Tommaso Zugno}
\affiliation{%
   \institution{University of Padova}
   \city{Padova}
   \country{Italy}}
\email{tommasozugno@gmail.com}

\author{Francesca Cuomo}
\affiliation{%
   \institution{University of Rome}
   \city{Rome}
   \country{Italy}}
\email{francesca.cuomo@uniroma1.it}

\author{Tommaso Melodia}
\affiliation{%
   \institution{Northeastern University}
   \city{Boston}
   \state{Massachusetts}
   \country{USA}}
\email{melodia@northeastern.edu}

\copyrightyear{2023}
\acmYear{2023}
\setcopyright{acmlicensed}\acmConference[WNS3 2023]{2023 Workshop on ns-3}{June 28--29, 2023}{Arlington, VA, USA}
\acmBooktitle{2023 Workshop on ns-3 (WNS3 2023), June 28--29, 2023, Arlington, VA, USA}
\acmPrice{15.00}
\acmDOI{10.1145/3592149.3592161}
\acmISBN{979-8-4007-0747-6/23/06}

\renewcommand{\shortauthors}{A. Lacava et al.}

\begin{abstract}

O-RAN is radically shifting how cellular networks are designed, deployed and optimized through network programmability, disaggregation, and virtualization. 
%
Specifically, \glspl{ric} can orchestrate and optimize the \gls{ran} operations, allowing fine-grained control over the network. \glspl{ric} provide new approaches and solutions for classical use cases such as on-demand traffic steering, anomaly detection, and \gls{qos} management, with an optimization that can target single \glspl{ue}, slices, cells, or entire base stations.
%
%
Such control can leverage data-driven approaches, which rely on the O-RAN open interfaces to combine large-scale collection of \gls{ran} \glspl{kpm} and state-of-the-art \gls{ml} routines executed in the \glspl{ric}. 
While this comes with the potential to enable intelligent, programmable \glspl{ran}, there are still significant challenges to be faced, primarily related to data collection at scale, development and testing of custom control logic for the \glspl{ric}, and availability of Open RAN simulation and experimental tools for the research and development communities.
To address this, we introduce ns-O-RAN, a software integration between a real-world near-real-time RIC and an ns-3 simulated RAN which provides a platform for researchers and telco operators to build, test and integrate xApps.
ns-O-RAN extends a popular Open RAN experimental framework (OpenRAN Gym) with simulation capabilities that enable the generation of realistic datasets without the need for experimental infrastructure.
We implement it as a new open-source ns-3 module that uses the E2 interface to connect different simulated 5G base stations with the RIC, enabling the exchange of E2 messages and RAN \glspl{kpm} to be consumed by standard xApps.
Furthermore, we test ns-O-RAN with the \gls{osc} and OpenRAN Gym \glspl{ric}, simplifying the onboarding from a test environment to production with real telecom hardware controlled without major reconfigurations required. ns-O-RAN is open source and publicly available, together with quick-start tutorials and documentation.
\end{abstract}

\glsresetall



\maketitle

\glsresetall
\section{Introduction}
\label{sec:intro}

Emerging wireless networking use cases (e.g., support for diverse traffic, private networks, and ultra-dense networks) are increasing the complexity of network deployments.
%
Traditional and monolithic \glspl{ran} architectures cannot support these demands, due to their inflexible, hardware-based designs, the need for on-site setup and labor, and the lack of support for customizable algorithmic optimization~\cite{condoluci2018softwarization}. 
%
The Open \gls{ran} paradigm is drastically changing the approach to the management and optimization of cellular networks, moving from static designs to more advanced and flexible network architectures, such as cloud-native and virtualized \glspl{ran}
that can better support the demands of \gls{5g} network use cases and provide improved network performance.





Specifically, 
the O-RAN Alliance
is implementing the Open RAN vision through technical specifications that extend the capabilities \gls{3gpp} networks. O-RAN networks are based on disaggregation, with network functions split across multiple software and white-box hardware components, virtualization, and programmability of the network. 
%
%
%
Algorithmic control is enabled by the \gls{ric}, a centralized abstraction of the network that has access to all the analytics collected by the \gls{ran} functions and can apply control actions. 
The \gls{ric} extends classical \gls{rrm} with data-driven approaches based on live telemetry from the \gls{ran} processed by \gls{ml} and \gls{ai} pipelines~\cite{bonati2021intelligence}.
These models are on-boarded in the software containers called xApps, which operate on a time scale between 10 ms and 1 s on the near-real-time RIC, and rApps, for control loops above 1 s in the non-real-time RIC~\cite{garcia2021ran}.

However, several challenges must be addressed to fully enable intelligent control in O-RAN, primarily in the area of datasets---that need to be representative and generalize well---and testing of the proposed solutions---which should not impact production network performance.
%
%
Indeed, O-RAN control needs to be effective at a large scale, and in generic scenarios, leveraging interactions and patterns that emerge when hundreds of \glspl{ue} interact with the network.
Such patterns are hard to reproduce because they require the use of expensive hardware, such as \glspl{sdr} or commercial equipment, or large-scale production deployments
to make sure that the data acquired can help the \gls{ai} model generalize well.
%
%
The \gls{sdr}-based framework OpenRAN Gym~\cite{bonati2022openrangym} has been designed to enable the end-to-end design of AI/ML pipelines for O-RAN and has shown very promising results for a variety of control use cases (e.g., slicing, neutral host applications), but still require dedicated hardware and experimental platforms.  
%
%
This practical limitation is usually addressed by the use of simulators, however, their integration with the O-RAN framework requires different development pipelines, meaning that moving a trained policy from the simulated environment to its real deployment as xApp requires additional efforts and thus costs~\cite{s21248173}.

%

To address such challenges, in this paper, we present ns-O-RAN, the first open-source simulation platform that combines a functional 4G/\gls{5g} protocol stack in ns-3 with an O-RAN-compliant E2 interface to enable the communication between the simulated environment and a near-real-time \gls{ric}.
ns-O-RAN extends the OpenRAN Gym framework with simulation capabilities in ns-3, enabling dataset generation for large-scale scenarios with hundreds of \glspl{ue}. ns-O-RAN is open source and publicly available as part of the \gls{osc} projects, together with tutorials and the additional Open RAN Gym components\footnote{\url{openrangym.com}}.

Through ns-O-RAN, we provide the capability to define simulated closed-loop control routines, where ns-O-RAN provides the cell \glspl{kpm} to the near-real-time \gls{ric}, and xApps in this platform define control actions which are sent back to the simulated environment. The latter then applies the control actions to the \gls{ran} model, adapting the behavior of the simulation according to the strategy defined by the O-RAN infrastructure. This enables the study of use cases that usually are related to large deployments such as \gls{ts} (e.g., to load balance users across cells), \gls{qos} (e.g., to control bearer parameters), and more. Specifically, we integrated ns-O-RAN with the base station implementation of the 5G module for ns-3 introduced in~\cite{mezzavilla2018end}.

ns-O-RAN also increases the versatility of the OpenRAN Gym approach, where the development of a xApp and \gls{ai}-based control policies can be first done on simulation, leveraging its capability of generating large-scale \gls{3gpp}-compliant datasets, and then transitioned to experimental platforms such as Colosseum for emulation with hardware in the loop~\cite{bonati2021colosseum}, and to over-the-air testbeds for further testing~\cite{bonati2022openrangym}. This can be done by exploiting a key characteristic of the ns-O-RAN design: we connect the emulated environment to any O-RAN-compliant near-real-time RIC with real xApps, which do not need to be re-implemented from scratch to work in the simulated and/or in the experimental environments. This is a deliberate design choice that simplifies the development life-cycle of end-to-end, intelligent control solutions for Open RAN. 

ns-O-RAN has been used in~\cite{lacava2022programmable} to demonstrate intelligent control of handovers in large-scale 5G scenarios. This paper provides a technical overview of how ns-O-RAN has been implemented, together with a tutorial on how it can be deployed with the OpenRAN Gym \gls{ric} and a profiling of the performance of the E2 interface.
The remainder of this paper is organized as follows.
In Section~\ref{sec:oran}, we present the O-RAN architecture and its main principles, describing what is the current state of the art for experimenting with this architecture.
In Section~\ref{sec:nsoran}, we introduce ns-O-RAN, discussing its open source components and the technical choices adopted to enable the end-to-end communication with the near-RT \gls{ric}.
In Sections~\ref{sec:tutorial} and~\ref{sec:results}, we present a practical example of an ns-O-RAN use case with a simple ns-3 scenario that is able to communicate with the near-real-time \gls{ric}, discussing also its preliminary results and showing the flow of the \glspl{kpm} from the simulated environment to the OpenRAN Gym near-real-time \gls{ric}~\cite{polese2022coloran}.
We conclude the paper and discuss what are the next steps in further extending ns-O-RAN in Section~\ref{sec:conclusions}.

\section{Toward Open RAN Networks}
\label{sec:oran}
The O-RAN architecture is a new approach to building mobile networks with greater innovation, competition, and interoperability in the telecommunications industry. The O-RAN architecture disrupts the classical approach by adopting the principles of {\em disaggregation}, {\em openness}, {\em virtualization}, and {\em programmability}, making it possible to expose data and analytics and enabling data-driven optimization, closed-loop control, and automation. In this section, we provide a brief introduction to the O-RAN architecture and then review the state of the art on tools for the design and development of O-RAN solutions.

\subsection{O-RAN: A Primer}
The \gls{ran} disaggregation splits base stations into different functional units: the \gls{ru}, the \gls{du} and the \gls{cu}.
The \gls{ru} manages \gls{rf} components and part of the physical layer, the \gls{du} provides support for the higher part of the physical layer, \gls{mac}, and \gls{rlc} layers, and the \gls{cu} features the higher layers of the protocol stack such as \gls{sdap}, \gls{pdcp} and \gls{rrc}.
This separation allows for the \gls{cu}, \gls{du} and \gls{ru} to be developed, procured and operated independently, enabling a more flexible and cost-effective network deployment~\cite{polese2022understanding}.
The interfaces between the different nodes are \textit{open} and standardized, to expose RAN telemetry and control to the external world, to achieve multi-vendor interoperability and the integration of different vendors' equipment and solutions into the network.

An additional core innovation is the \gls{ric}, a new architectural component that provides a centralized abstraction of the network, allowing operators to implement and deploy custom control plane functions for the integration of new technologies, such as \gls{5g} and \gls{ai}, into the network.
O-RAN envisions different loops operating at timescales that perform management and control of the network at near-real-time from 10 ms to 1 s, through third-party applications called xApps, and non-real-time, i.e., for more than 1 s through applications called rApps.

The near-RT RIC is the core of the control and optimization of the RAN and its main components are the xApps.
A xApp is a plug-and-play element that implements custom logic, for example for RAN data analysis and RAN control. xApps can receive data and telemetry from the RAN and send back control using the E2 interface.
The E2 interface is an open interface between two endpoints, i.e., the near-RT RIC and the so-called E2 nodes, i.e., DUs, CUs, and O-RAN-compliant LTE eNBs. The E2 interface has been logically structured into two distinct protocols: the \gls{e2ap} and \gls{e2sm}. The \gls{e2ap} is a fundamental procedural protocol that facilitates coordination between the near-RT RIC and the E2 nodes, dictating how they communicate with one another and providing a basic set of services. \gls{e2ap} messages can embed different \gls{e2sm}s, which
implement specific functionalities (i.e., the reporting of RAN metrics or the control of RAN parameters).

\subsection{Experiments and Simulations for O-RAN}

O-RAN development is still in its early stage, thus there is a limited number of solutions for developing and testing on the O-RAN architecture. 

Two frameworks were recently developed to overcome several problems related to dataset availability, developing, designing, prototyping, and testing O-RAN-ready solutions. 
OpenRAN Gym~\cite{bonati2022openrangym}, an open-source toolbox to develop \gls{ai}/ML O-RAN-compliant inference and control algorithms, to deploy them as xApps on the near-RT \gls{ric}, and to test them on a large-scale softwarized \gls{ran} controlled by the \gls{ric}. OpenRAN Gym is platform-independent, and it allows users to perform data collection campaigns to build datasets, develop, design, prototype, and test O-RAN-ready solutions at scale.
This framework provides a lightweight implementation of the \gls{osc} near-RT \gls{ric} which has been adapted to run on the Colosseum system as a set of standalone Docker containers—as well as automated pipelines for the deployment of the various services of the \gls{ric}. OpenRAN Gym leverages srsRAN and OpenAirInterface to implement the \gls{ran} through software-defined radios.
The Open AI Cellular (OAIC) initiative has introduced an O-RAN framework to manage cellular networks---based on srsRAN---through \gls{ai}-enabled controllers, and to interact with systems that locate implementation, system-level, and security flaws in the network itself~\cite{upadhyaya2022prototyping}.

In this paper, we extend the OpenRAN Gym with an ns-3 integration, to provide an environment that does not require experimental infrastructure and software-defined radios.
\gls{ns3} is a discrete-event network simulator which is highly flexible and customizable: users can configure various simulation parameters, such as network topology, node mobility, and traffic patterns, to match the specifics of their network scenario. 
Specifically, we leverage the availability of very accurate \gls{3gpp} stochastic models~\cite{zugno2020implementation} 
and the \gls{ns3} 5G module from~\cite{mezzavilla2018end}. This extends the \gls{ns3} LTE module with additional features such as 5G-compliant physical and \gls{mac} layers, antenna patterns, beamforming algorithms, and mobility procedures for multi-connected devices.

The \gls{ns3} simulator is also an ideal platform for enhancing \gls{ai} solutions for networks.
In recent years, various studies have extended the standard capabilities of \gls{ns3} by integrating its potential with popular machine learning development software. This has allowed for the creation of more advanced \gls{ai} solutions for networks.
In~\cite{ns3gym}, authors propose \gls{ns3}-gym, a framework that integrates OpenAI Gym and \gls{ns3}, two popular tools in the fields of reinforcement learning and network simulation, respectively. ns3-gym enables the use of reinforcement learning techniques for network optimization and management problems, by combining the simulation capabilities of \gls{ns3} with the reinforcement learning algorithms of OpenAI Gym. 
\gls{ns3}-ai~\cite{ns3ai} provides a high-efficiency solution for data interaction between \gls{ns3} and other python-based \gls{ai} frameworks, following the principles of \gls{ns3}-gym.
However, neither \gls{ns3}-gym nor \gls{ns3}-ai can be used as a framework for developing O-RAN xApps that are suitable for immediate use in a production environment, unlike the proposed ns-O-RAN framework in this paper, which we describe in details next.




\section{ns-O-RAN}
\label{sec:nsoran}

\begin{figure}[t]
    \centering
    \includegraphics[width=.95\columnwidth]{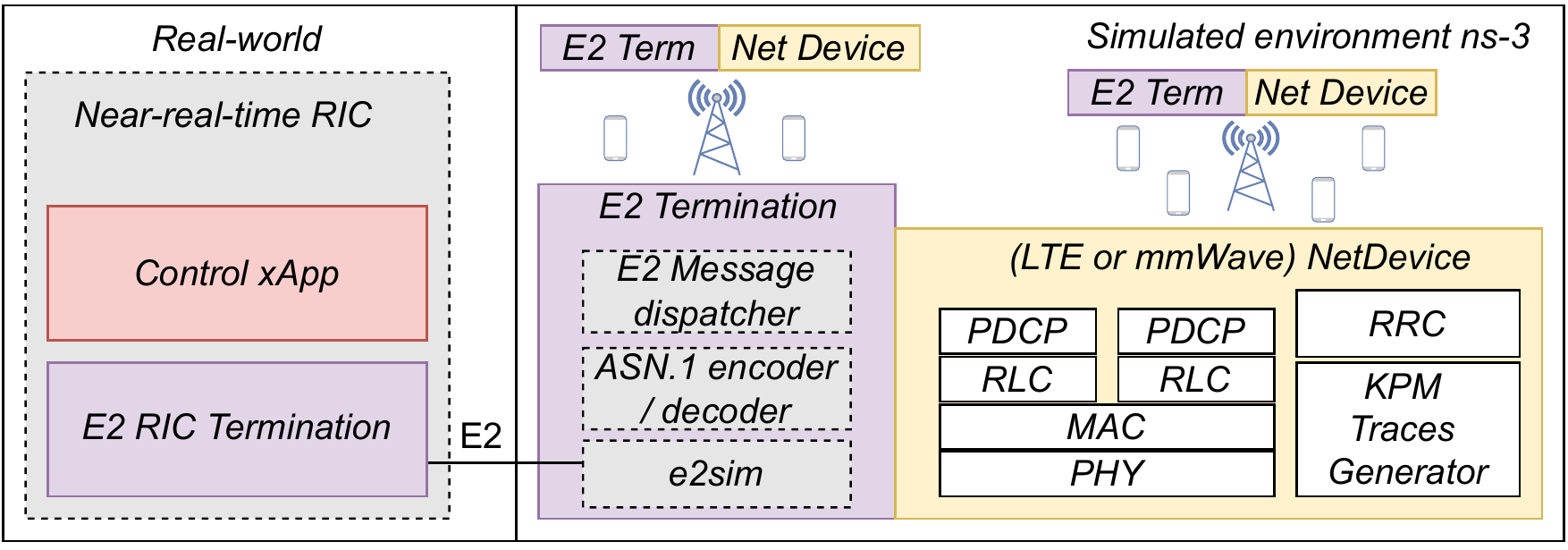}
    \caption{ns-O-RAN Architecture}
    \label{fig:architecture}
\end{figure}

ns-O-RAN extends the OpenRAN Gym framework, embedding the first open-source simulation platform that combines a functional 4G/5G protocol stack in ns-3 and an O-RAN-compliant E2 interface for simulated base stations.
Such a platform was designed to enhance data collection and xApp testing capabilities, a critical step toward enabling efficient and generic \gls{ai} and \gls{ml} solutions for Open \gls{ran} and the next generation of cellular network systems.
To this end, ns-O-RAN enables the integration of O-RAN software such as the OpenRAN Gym and \gls{osc} near-RT \glspl{ric} with large-scale \gls{5g} simulations using \gls{3gpp} channel models and detailed modeling of the full \gls{3gpp} \gls{ran} protocol stack.
This allows data collection of \gls{ran} \glspl{kpm} at scale, in different simulated scenarios, and with different applications (e.g., multimedia streaming, web browsing, wireless virtual reality, etc). 

ns-O-RAN supports an O-RAN-compliant E2 interface and implements two different \glspl{e2sm}, i.e., \gls{e2sm} \gls{kpm}, for reporting, and \gls{e2sm} \gls{rc}, to push control actions to the \gls{ran} (for example, of traffic steering and mobility management).
At its core, ns-O-RAN is an ns-3 external module that provides an \gls{sctp} connection between the simulator and the near-RT \gls{ric} for the support of the \gls{e2ap} and \gls{e2sm}.
The E2 termination of the \gls{ric} can connect to a set of E2 terminations running in the ns-3 simulation, responsible for handling all the E2 messages from and to the simulated environment.
This connection was developed by extending the \gls{osc} E2 simulator (i.e., \texttt{e2sim}) and wrapping it into an ad hoc module for ns-3.

This framework can generate realistic datasets based on stochastic \gls{3gpp}-defined wireless channel, that can be fed straight into the O-RAN \gls{ric} with no need for infrastructure.
Moreover, ns-O-RAN inherits the ease of customization from ns-3, i.e., it can be easily configured to support countless possible scenarios and use cases that can be studied over the same platform.
Such scenarios can then be the foundation to build O-RAN-compliant xApps that can be tested and tuned on ns-3 and then deployed on a real \gls{ran} with no software changes.

As shown in Figure~\ref{fig:architecture}, ns-O-RAN features three different software applications: (i) the \texttt{e2sim}~\cite{e2sim} software, which was originally developed by the \gls{osc} community and extended as part of this effort to process \gls{e2sm} RC handover management actions and handle multiple E2 terminations; (ii) the ns3-mmWave module~\cite{mezzavilla2018end}; and (iii) the ns-O-RAN module, introduced in~\cite{lacava2022programmable}, which is an ns-3 external module that uses the \texttt{e2sim} to create a \gls{sctp} connection with the \gls{ric} and provides E2 terminations to the simulated base stations.

\subsection{\texttt{e2sim}}
\label{sec:e2sim}
The E2 simulator, namely \texttt{e2sim}~\cite{e2sim}, is an \gls{osc} software designed to simulate the \gls{ran} E2 termination to allow the development of the hosts on near-RT \gls{ric} and xApps.
It is an \gls{sctp} client that implements the \gls{e2ap} basic specifications, allowing for end-to-end E2 flow testing by creating a connection to the E2 Termination in a near-RT \gls{ric} platform. 
It is capable of decoding incoming messages from the \gls{ric} and providing feedback, as well as streaming \gls{ran} telemetry to the \gls{ric}. 
The system also behaves as an external interface library, exposing a main class called \texttt{E2Sim} that provides basic APIs for managing connection and message reception.
By referencing this class in their codebase, external applications can establish a direct connection with the \gls{ric}.



More precisely, at startup, the \texttt{e2sim} generates one E2 Setup Request for each \gls{ran} node, including the \gls{ran} function definition and identifiers (IDs) that specify the \gls{ran} node capabilities (e.g., control actions that it can ingest, performance metrics it can report).
The \gls{ric} replies with an E2 Setup Response for each request, to acknowledge the presence of the nodes within the network. \texttt{e2sim} uses an event-driven system made of callbacks to forward the messages from the \gls{ric} to the external calling applications.

When a new message from the \gls{ric} is received, \texttt{e2sim} decodes it and identifies the \gls{ran} Function ID to trigger an event to all the registered callbacks sharing the incoming message.
If no callback is registered, no action is taken.
The calling application can subscribe to new callbacks using the \gls{e2sm} \gls{ran} Function ID in \texttt{e2sim} to notify its capability to support a specific feature.
The xApp can register to the \glspl{e2sm} by sending a \gls{ric} Subscription Request indicating which \gls{ran} function is going to use during its working.
It will be then the responsibility of the calling application to generate and send the appropriate responses to the xApp and this can be done by using the E2Sim \texttt{SendMessage} function.




{\bf Extension of the \texttt{e2sim} library --}
The original \texttt{e2sim} project was designed to simulate the E2 \gls{ran} Termination of a single \gls{gnb} that is connected to the near-RT \gls{ric}.
Therefore its software base relies on this design choice and uses global variables that are meant to describe the features of the \gls{bs}, such as the \gls{gnb} identifier, the \gls{sctp} socket file descriptor, and the client port  that determine the connection in the Linux system.
On the contrary, in ns-3 all the \glspl{bs} are managed by a single process that simulates the interactions in form of time-based scheduled events.
For this reason, we extended the \texttt{e2sim} library\footnote{https://github.com/wineslab/ns-o-ran-e2-sim} to support multiple \glspl{gnb} simultaneously over the same machine. To do so, we removed the global variables and integrated them as configurable parameters of the \texttt{E2Sim} class.
Moreover, compared to a standard \gls{ran} deployment, where the \glspl{bs} are expected to have different IP addresses that distinguish the connections to the \gls{ric}, ns-3 runs on a single process/host, making it more difficult to expose different IPs.
To address this issue and improve the performance of the simulation, we developed a novel approach that uses multiple threads in ns-3 to establish independent \gls{sctp} connections for each E2 link. 
Specifically, we created new parameters to support the inclusion of a local port number and extended ns-3 to create separate threads guaranteeing the independence of each E2 data flow.

By doing so, we can successfully establish connectivity between multiple \gls{ran} nodes and the near-RT \gls{ric} even if a single IP is associated with the simulation process.

Finally, in the original \texttt{e2sim} codebase the only \gls{ran} function ID supported was the number 200, which identifies the possibility for the \gls{bs} to collect \gls{kpm} reports and send them to the \gls{ric} through E2 Indication Messages.
In our version, we extended the \gls{e2sm} capabilities including the \gls{rc} service model, which allows the xApp to send to the \gls{ran} control actions.
To support these extensions, we have integrated the callback system in our application to handle the \gls{rc} messages from the xApp, thus enabling the \gls{ran} control.

\subsection{The ns-O-RAN Module}
  \begin{figure*}[t]
    \centering
    \includegraphics[scale=0.5]{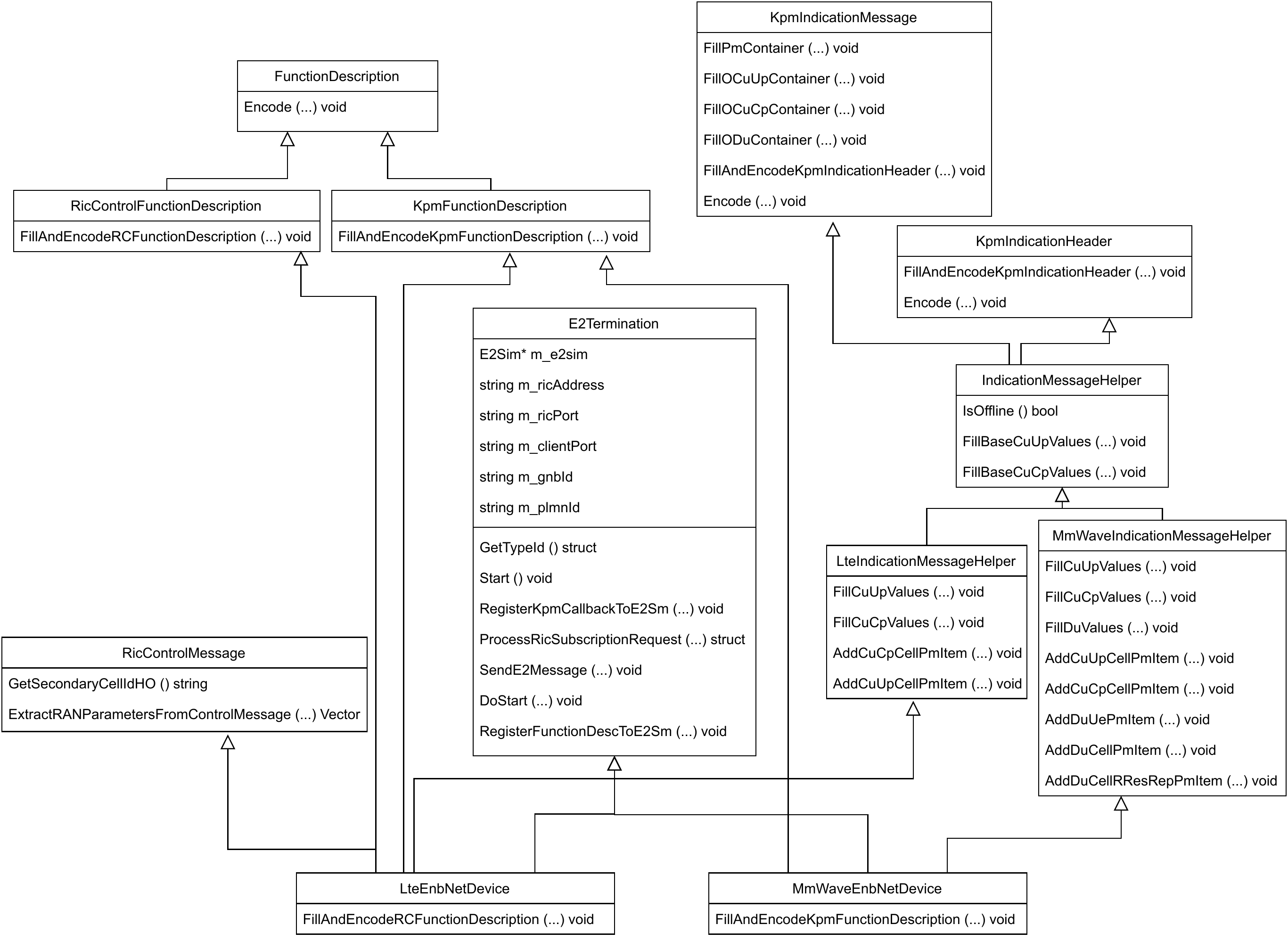}
    \caption{Simplified UML Diagram of ns-O-RAN and its Main Connections with the ns-3 mmWave Module}
    \label{fig:uml}
\end{figure*}

The integration between ns-3 and \texttt{e2sim} has been implemented through a dedicated
module\footnote{https://github.com/o-ran-sc/sim-ns3-o-ran-e2}.
%
It uses \texttt{e2sim} as a library and implement its callbacks, besides providing streamlined support for multiple E2 termination instances configuration.
These operations are handled by the \texttt{E2termination} class, which wraps all the major \texttt{e2sim} functionalities exposing them to ns-3.
Figure~\ref{fig:uml} shows the simplified \gls{uml} diagram, illustrating how the ns-O-RAN classes interact with each other and how they work in connection with the ns-3 mmWave module.
To enable the exchange of the \gls{ran} functions and capabilities through \gls{e2ap}, we create the O-RAN \gls{ran} Function descriptors in the simulation with the auxiliary class \texttt{FunctionDescription} and its extensions \texttt{RicControlFunctionDescription} and \texttt{KpmFunctionDescription}. 

%

The ns-O-RAN module implements the creation and encoding of periodic reports from the \gls{ran} using the \gls{ric} Indication Messages according to the O-RAN \gls{kpm} specification~\cite{oran-wg1-use-cases-analysis}.
This is implemented in the \texttt{KpmIndicationHeader} and in the \texttt{KpmIndicationMessage} classes that are created and encoded using the \texttt{IndicationMessage\-Helper}.
Moreover, we developed the \gls{rc} capability for the O-RAN \gls{ts} use-case~\cite{oran-wg3-use-cases} through the Handover Management message, so that the xApp can send a \gls{rc} Control Message with \gls{ran} Function ID equal to 300. 
In this message, the xApp specifies the identity of an \gls{ue} and the target cell to which the handover should be performed.
The procedure is initiated by a source \gls{gnb}, which transfers the \gls{ue}'s context to the target cell.
The \gls{ric} Control Messages are decoded in the ns-O-RAN module using the class \texttt{RicControlMessage}.
Finally, the \texttt{asn1c-types} files contain wrapper classes for the \gls{asn1} C structures that have re-designed to improve their allocation and deallocation and thus the general memory management of the module.

The ns-O-RAN module follows the classical installation steps of an external contribution module of ns-3
and requires the \texttt{e2sim} library to be installed to build the whole ns-3 project.

\subsection{Integration with the 5G Cellular Model}

The modules described in the previous sections are responsible for the setup of the connection between the simulated \gls{ran} and the \gls{ric} through the E2 interface.
Such modules are designed to be agnostic with respect to the ns-3 module in use and they can be integrated with any cellular network implementation done for ns-3.
For this work, we extended the ns-3 mmWave module\footnote{https://github.com/wineslab/ns-o-ran-ns3-mmwave} for the simulation of 5G cellular networks, providing its simulated base stations the E2 Termination support.

The ns-3 mmWave module~\cite{mezzavilla2018end} is designed to perform end-to-end simulations of \gls{3gpp}-style cellular networks.
Built upon the ns-3 \gls{lte} module (LENA)~\cite{lena}, this 5G module enables the support of a wide range of channel models for frequencies between 0.5 and 100 GHz, thus including 3GPP NR \gls{fr1} and \gls{fr2}.
The \gls{phy} and \gls{mac} classes in this system have been specifically designed to support the \gls{3gpp} NR frame structure and numerologies, ensuring compatibility with the latest cellular technologies.
At the \gls{mac} layer, it supports carrier aggregation~\cite{zugno2018integration} and multiple scheduling policies, to provide additional capacity. 
Finally, the module also enables dual connectivity with the \gls{lte} base stations, featuring fast secondary cell handover and channel tracking.
For a full list of the features implemented in this module, we refer the reader to the original repository of the project\footnote{https://github.com/nyuwireless-unipd/ns3-mmwave}.

\begin{figure}[b]
    \includegraphics[width=1\columnwidth]{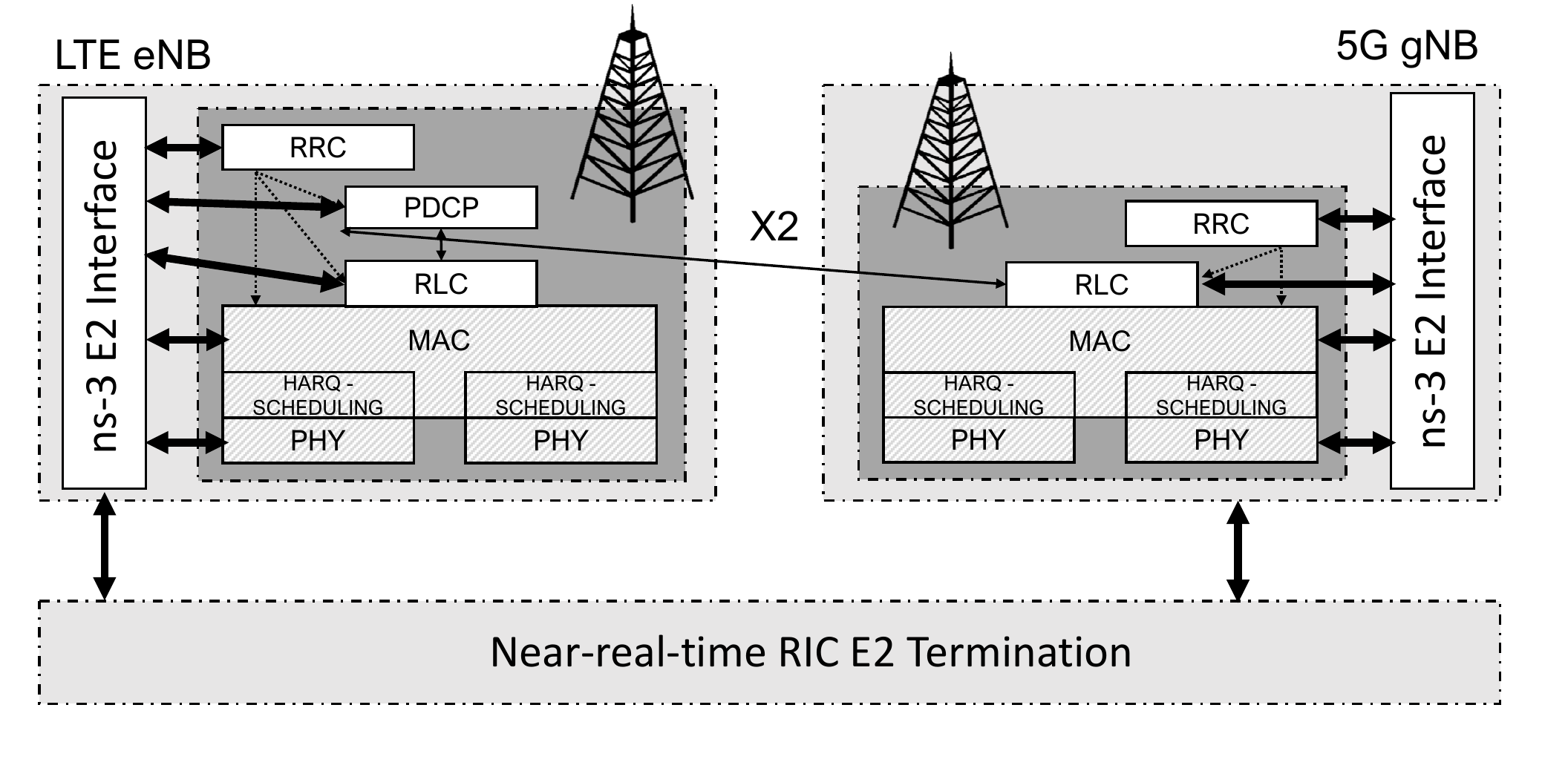}
    \caption{Integration Between the near-RT \gls{ric} and the ns-3 mmWave Models Through ns-O-RAN}
    \label{fig:dual_conn}
\end{figure}

The goal of this integration is to enable an end-to-end \gls{ran} simulation for the O-RAN \glspl{ric}. 
This requires the \gls{ran} side 
to enable \gls{e2sm} exchanges with the \gls{ric}, i.e., the delivery of the \gls{kpm} reports and the digestion of the control actions. The architecture of the ns-3 mmWave module along with the structural changes we performed to integrate ns-O-RAN with it are shown in Figure~\ref{fig:dual_conn}.

In the original module, the \gls{ran} functions are mainly managed by two central \texttt{NetDevice} classes called \texttt{LteEnbNetDevice} and \texttt{MmWaveEnbNetDevice}, which are responsible to generate and trigger the functions for the management of the full radio protocol stack and of the \gls{rlc}, \gls{rrc} and \gls{pdcp} data flows in the simulation.
The first major design modification was to adapt these classes to support the data collection according to the O-RAN specifications.
%
%
We created three new functions also shown in the simplified \gls{uml} of Figure~\ref{fig:uml}, which are responsible for the generation of the \glspl{kpm} from the simulation and their dispatch through the E2 Termination.
Each function manages the data relative to one of the O-RAN specified disaggregated units, namely \gls{cucp}, \gls{cuup}, and \gls{du}. 
Each disaggregated unit has its own statistics calculator that is able to generate and send to the \gls{ric} a different Indication Message.
At the current moment, we do not support the \gls{lte} \gls{du} reports. 
Both the NetDevices are registered to the E2 updates through the callback system described in the previous sections for the \gls{kpm} \gls{ran} Function description.

With respect to the control, the mmWave module implements a \gls{nsa} architecture having a Primary \gls{lte} cell, also called \gls{enb}, that is responsible for the \gls{pdcp} operations of the \gls{ran} and set of secondary \gls{gnb} 5G cells.
This led us to implement the processing of the control actions only in the \texttt{LteEnbNetDevice} class.

Following the O-RAN specifications, we adapted the code of the base station \texttt{NetDevices} to include an object of type \texttt{E2Termination} as the \texttt{m\_e2term} private attribute to  provides E2 connectivity through ns-O-RAN and \texttt{e2sim}.
In this way, when the ns-3 simulation starts, every \texttt{LteEnbNetDevice} and every \texttt{MmWaveEnbNetDevice} instantiated creates its own instance of an \texttt{E2Termination} C++ object to represent different \gls{ran}-side E2 terminations that implement the message exchange described in Section~\ref{sec:e2sim}.
This operation is done by the \texttt{MmWaveHelper} class, which is also responsible for instantiating the objects used to trace and compute the \gls{pdcp}, \gls{rlc} and the \gls{mac} and physical layer measurement that the E2 terminations will send to the near-RT \gls{ric}.
This process is transparent to the \gls{ric}, which indeed does not need to be aware that the \gls{ran} it is serving is a simulated one.
Each \gls{ran} function is bound to just one E2 interface, as depicted in Figure~\ref{fig:architecture}, and has its own socket pair address.

One of the challenges of integrating simulating systems with software running in the real world is time synchronization among the different hosts.
While the real-world near-RT \gls{ric} is expected to work with control loops between 10 ms and 1 s, \gls{ns3} is a discrete-event framework that can execute faster or slower than the wall clock time, thus generating a possible time gap between the two systems.
To synchronize the two systems, at the beginning of the simulation \gls{ns3} stores the current Unix time in milliseconds and uses it as the baseline timestamp.
Whenever an E2 message is sent to the \gls{ric}, \gls{ns3} will sum the simulation time elapsed and the baseline timestamp.
In this way, the \gls{ric} is always able to correctly reorder the messages preventing any inconsistency and invalidation of the received data.
The xApps onboarded on the \gls{ric} shall use the timestamp provided by these messages to follow the simulation and eventually generate control actions based on it.


We refer to the interactions between the 5G module and the near-RT \gls{ric} described in this section as \textit{online mode} of ns-O-RAN, since the absence of an \gls{sctp} connection with the \gls{ric} would prevent the correct behavior and collection of the simulation data.
Nevertheless, the ns-O-RAN and ns-3 mmWave modules can also work in a \textit{stand-alone} or \textit{offline mode}, where at the beginning of the simulation no connection is established and the \glspl{kpm} collected by the three main functions are stored as traces on files in the local machine.
This mode is especially useful for large data collection campaigns (e.g., orchestrated with the Simulation Execution Manager~\cite{magrin2019sem}), where the main goal is not to control the simulation flow, but to collect data to later study the \gls{ran} behavior.

\section{End-to-end Deployment with OpenRAN Gym RIC}
\label{sec:tutorial}

In this section, we show how to set up ns-O-RAN on a virtualized environment to create a connection between the \gls{ric} and ns-3 and to allow the exchange of the E2 messages, thus creating a \textit{simulated} closed control loops between ns-3 and a near-RT \gls{ric}.
For this tutorial, we use the near-RT \gls{ric}\footnote{\url{https://github.com/wineslab/colosseum-near-rt-ric/tree/ns-o-ran}} from the OpenRAN Gym framework~\cite{polese2022coloran}, which can be installed on a local workstation or loaded into experimental platforms such as Colosseum~\cite{bonati2021colosseum}.
In this latter case, we already provide a publicly available Colosseum LXC image.

From the \gls{ran} side to the \gls{ric}, we can collect the implemented \glspl{kpm} from the simulation, wrap them up into \gls{ric} Indication Messages and send them through the E2 interface.
The \gls{ric} is then able to receive such messages, read the \glspl{kpm} and \glspl{kpi} to define a data-driven policy (or to simply apply an adaptive policy), create a Control Action and then send the action inside a \gls{ric} Control Message to implement the policy.
Finally, the \gls{ran} receives the Control Message and applies the changes request by the \gls{ric}.
We define these exchanges as a \textit{simulated control loops} since the control actions sent to the \gls{ric} can tune the simulation and change its action in a controlled and reproducible environment.

\subsection{Local Setup of the near-RT RIC}
A near-RT \gls{ric} (e.g., the OpenRAN Gym \gls{ric}) is required to work with ns-O-RAN in its online mode.
First, the \gls{ric} repository shall be cloned and set up by configuring and executing the main components that are going to interact with the ns-3 environment:
\begin{itemize}
    \item the near-RT \gls{ric}, which includes multiple Docker containers for the E2 termination, an internal message routing manager, a database, and a manager for the E2 connections; and
    \item the sample xApp that will receive  and process the indication messages from ns-O-RAN.
\end{itemize}
These entities can be instantiated by running the commands provided in Listing~\ref{lst:ric-setup}.

\begin{lstlisting}[language=mybash,style=mystyle,
caption={Commands to Start the near-RT \gls{ric}},
label={lst:ric-setup}]
#!/bin/bash
git clone -b ns-o-ran https://github.com/wineslab/colosseum-near-rt-ric
cd colosseum-near-rt-ric/setup-scripts
./import-wines-images.sh  # import and tag base images from Docker hub
./setup-ric-bronze.sh  # setup and launch
\end{lstlisting}

Once this is done the \gls{ric} components should be up and running in different Docker containers (which can be listed using the \texttt{docker ps} command).
As a next step, we advise opening a terminal window for logging the values of the E2 Termination and check the E2AP messages exchange, with the command shown in Listings~\ref{lst:e2term}. 
This helps understanding if \gls{e2ap} and \gls{e2sm} messages are received correctly.

\begin{lstlisting}[language=mybash,style=mystyle,
caption={Commands to Show \gls{ric} E2 Termination Logs. The Last \texttt{grep} Command Will Filter the Output to Show only when a Base Station is Interacting},
label={lst:e2term}]
#!/bin/bash
docker logs e2term -f --since=1s 2>&1 | grep gnb: 
\end{lstlisting}

\subsection{Installation and Connection with ns-O-RAN}
Several options are available to install ns-3 and the ns-O-RAN extension,
including the use of a \textit{Dockerfile} provided in the root of the \texttt{ns-o-ran} branch of the OpenRAN Gym near-RT RIC repository discussed above.
The three components required by ns-O-RAN need to be installed in sequence to properly configure the system, as shown in Listing~\ref{lst:ns-o-ran-setup}.

At first, the enhanced version of the \texttt{e2sim} should be cloned, built and installed with all the due prerequisites.
In this step, it is possible to set the verbosity of the \texttt{e2sim} by changing the relative argument passed to the build script.
This software is a strict dependency for the ns3-mmWave version adapted for O-RAN, and thus it should be set up before installing the main toolchain.
After this step, it is possible to set up the ns-3 mmWave main project and ns-O-RAN module, which is basically an external module that can be plugged in ns-3 and uses the \texttt{e2sim} to create a \gls{sctp} connection with the RIC.
Finally, we can clone the ns-O-RAN module and add it in the ns3-mmWave project in the \texttt{contrib} directory and we can build ns-3.

\begin{lstlisting}[language=mybash,style=mystyle,
caption={Commands to Setup ns-O-RAN Components and Build the ns-3 Toolchain},
label={lst:ns-o-ran-setup}]
#!/bin/bash
cd oran-e2sim/e2sim/
mkdir build
./build_e2sim.sh <verbosity> # verbosity = [0,3]
cd ../../
# clone mmwave project in a folder called ns-3-mmwave-oran
git clone https://github.com/wineslab/ns-o-ran-ns3-mmwave ns-3-mmwave-oran
cd ns-3-mmwave-oran/contrib
# clone ns-O-RAN from the OSC repository in a folder called oran-interface
git clone -b master https://github.com/o-ran-sc/sim-ns3-o-ran-e2 oran-interface
cd ..  # go back to the ns-3-mmwave-oran folder

./waf configure --enable-examples --enable-tests
./waf build
\end{lstlisting}

\subsection{Simulating the RAN --- Sample Scenario}

\begin{figure}[t]
    \centering
    \includegraphics[width=\columnwidth]{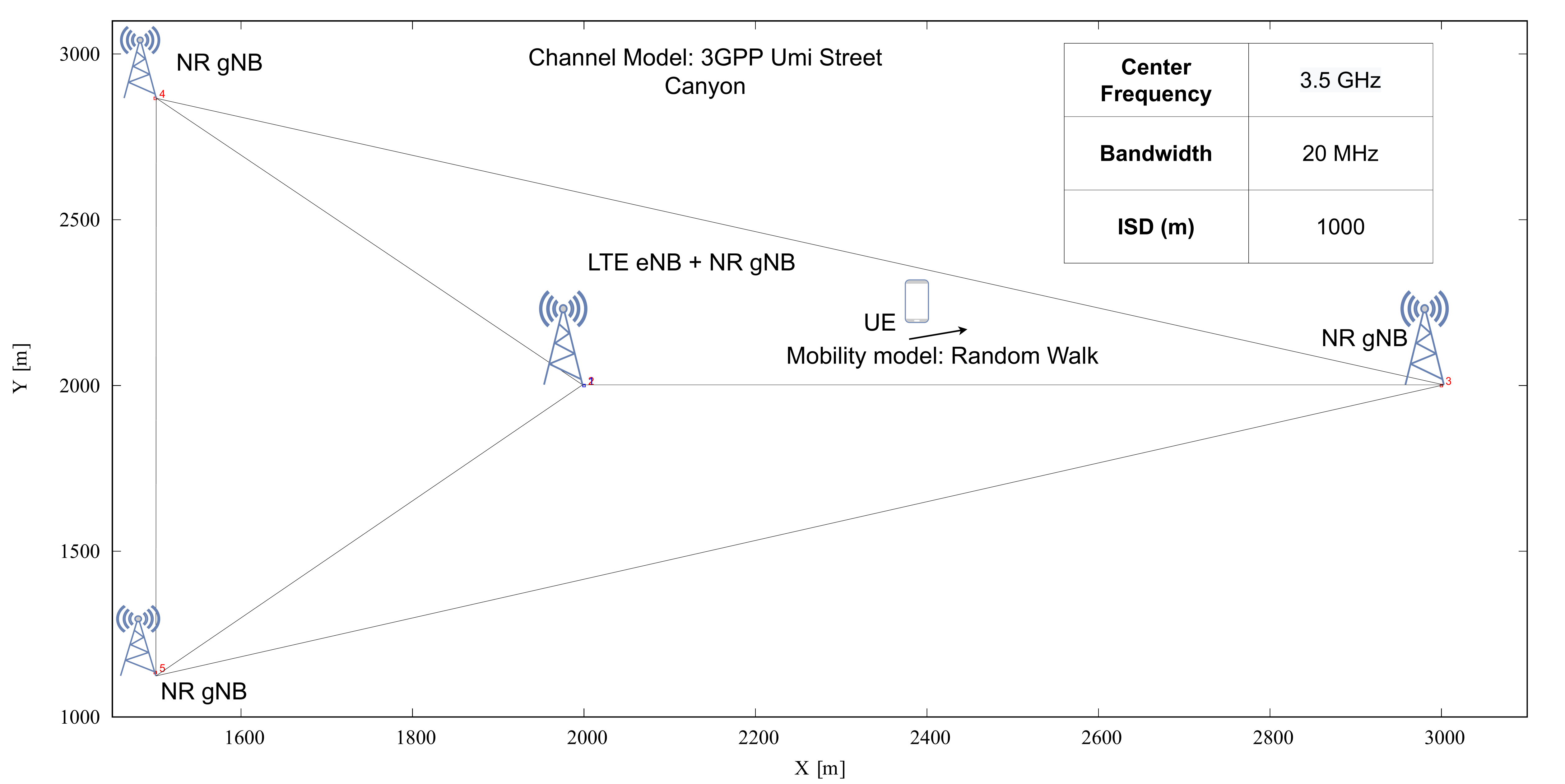}
    \caption{Scenario Zero Implemented to Test the Simulated Control Loop on ns-O-RAN}
    \label{fig:scenariozero}
\end{figure}

To instantiate the \gls{ran} and test connectivity to the \gls{ric}, we provide a sample ns-3 scenario in the file \texttt{scenario-zero.cc}. 
As shown in Figure~\ref{fig:scenariozero}, this scenario features a \gls{nsa} \gls{5g} setup with one \gls{lte} \gls{enb} in the center of the scenario, one \gls{gnb} co-located with the \gls{lte} base station and three \glspl{gnb} around it with an inter-site distance of 1000 meters with the center of the scenario.

By default, this scenario is deployed on the \gls{fr1} frequency range, using as center frequency 3.5 GHz with a bandwidth of 20 MHz.
One antenna is assigned to each \gls{ue} and \gls{bs}.
This scenario is a simplification of the Dense Urban Scenario described in the \gls{3gpp} Technical Release 38.913~\cite{3gpp.38.913}.
The channel model is the 3GPP UMi-Street Canyon from~\cite{3gpp.38.901}.

The number of \glspl{ue} in this scenario is 12 and they are allocated in their initial positions uniformly, i.e., with constant density, randomly within a disc having as center the co-located \gls{enb}-\gls{gnb} stations and a radius equal to the inter-site distance.
The mobility model of the \glspl{ue} is a random bi-dimensional walk. Each \gls{ue} moves with a uniform speed between 2 m/s and 4 m/s.
The source traffic model simulates \gls{embb} \glspl{ue} with a full buffer constant transmission of 20.48 Mbps.

The simulation time of this scenario is 2 seconds and the periodicity of the generation and delivery to the \gls{ric} of the \gls{e2sm} messages is 100 ms.

We can run the scenario by using the commands in Listing~\ref{lst:scen-zero}.
ns-O-RAN is designed to be compliant with a generic O-RAN-compliant \gls{ric}, thus the simulation scenario only needs the RIC E2 termination IP address, which can be configured using the \texttt{e2TermIp} attribute.


\begin{lstlisting}[language=mybash,style=mystyle,
caption={Commands to Run the Scenario Zero},
label={lst:scen-zero}]
#!/bin/bash
./waf --run scratch/scenario-zero.cc --e2TermIp="10.0.2.10"
\end{lstlisting}

\subsection{xApp Instantiation}

After the RAN is connected to the \gls{ric}, it is possible to start the xApp and log its container to verify that it receives E2 messages.
The procedure to start the sample xApp we provide as a Docker container is shown in Listing~\ref{lst:xapp-setup}.

\begin{lstlisting}[language=mybash,style=mystyle,
caption={Commands to Create the xApp Container and Run the xApp Logic},
label={lst:xapp-setup}]
#!/bin/bash
cd colosseum-near-rt-ric/setup-scripts
./start-xapp-ns-o-ran.sh
# In Docker
cd /home/sample-xapp
./run_xapp.sh
\end{lstlisting}

\section{Example Results}
\label{sec:results}
 

 \begin{figure}[hb]
    \centering
    \includegraphics[width=1.05\columnwidth]{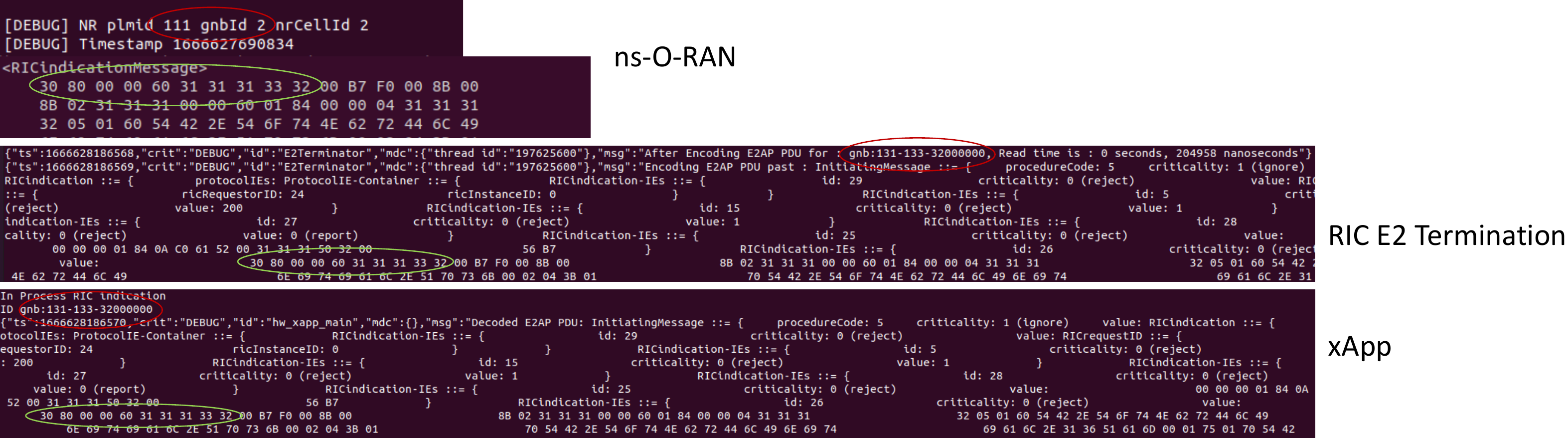}
    \caption{\gls{ric} Indication Message Read on ns-O-RAN, the E2 Termination and the xApp}
    \label{fig:tutorial}
\end{figure}

\begin{table*}
  \caption{\gls{e2ap} Traffic Exchange Between ns-O-RAN and the xApp During the Execution of a Simulation with the Sample ns-O-RAN Scenario.
  No Control Message Was Sent by the \gls{ric} During this Experiment}
\centering
  \begin{tabular}{llll}
    \toprule
    {\bf Measurement} & {\bf All} & {\bf Single eNB} & {\bf Single gNB}\\ \midrule
   {\bf Wireshark Filter} & \texttt{e2ap} & \texttt{e2ap and sctp.port == 38471} & \texttt{e2ap and sctp.port == 38472} \\ \midrule
   {\bf Number of Packets}   &  157  &  40 & 40 \\
   {\bf Time span (s)} & 439.220 & 429.324 & 439.211 \\
   {\bf Average pps} & 0.4 & 0.1 & 0.1 \\
   {\bf Average. size (B)} & 396 & 259 & 661 \\
   {\bf Bytes exchanged} & 62148  & 10352 & 26456 \\
   {\bf Average Data Rate (Bps)} & 141 & 24 & 60 \\
   {\bf Average Data Rate (bps)} & 1.131 & 192 & 481 \\
    \bottomrule
\end{tabular}
  \label{tab:recapzero}
\end{table*}

By running the sample scenario, it is possible to observe the message exchange between the simulated \gls{ran} and the real-world \gls{ric}:

\begin{itemize}
    \item E2 Setup Request (from ns-O-RAN to E2 Termination on the RIC);
    \item E2 Setup Response (from the E2 Termination on the RIC to ns-O-RAN);
    \item E2 Subscription Request (from the xApp to ns-O-RAN through the E2 Termination on RIC, after the xApp is instantiated);
    \item E2 Subscription Response (from ns-O-RAN to the xApp through the E2 Termination on RIC);
    \item E2SM RIC Indication Message (from ns-O-RAN to the xApp through the E2 Termination on RIC).
\end{itemize}
There are several ways to analyze and study these messages.
If the \texttt{verbosity} parameter has been set to 3 during the \texttt{e2sim} installation, ns-3 will show the encoded \gls{e2ap} messages that are generated and sent to the \gls{ric}.
The same Indication messages are also logged in the E2 Termination of the \gls{ric} and in the xApp, which decodes them, extracts the \glspl{kpm}, and digest them according to xApp program logic. Figure~\ref{fig:tutorial} shows the logs of three different components involved in the communication. 
In each of these logs, it is possible to inspect the Indication Messages, thus confirming the correct behavior of the end-to-end RAN-to-xApp communication.
We show in red the \gls{gnb} ID and in green the initial header of the message.
The \gls{gnb} ID format for ns-O-RAN on the near-RT \gls{ric} is gnb:131-133-3<gnbId><padding>

\begin{figure}[h]
    \centering
    \includegraphics[width=1\columnwidth]{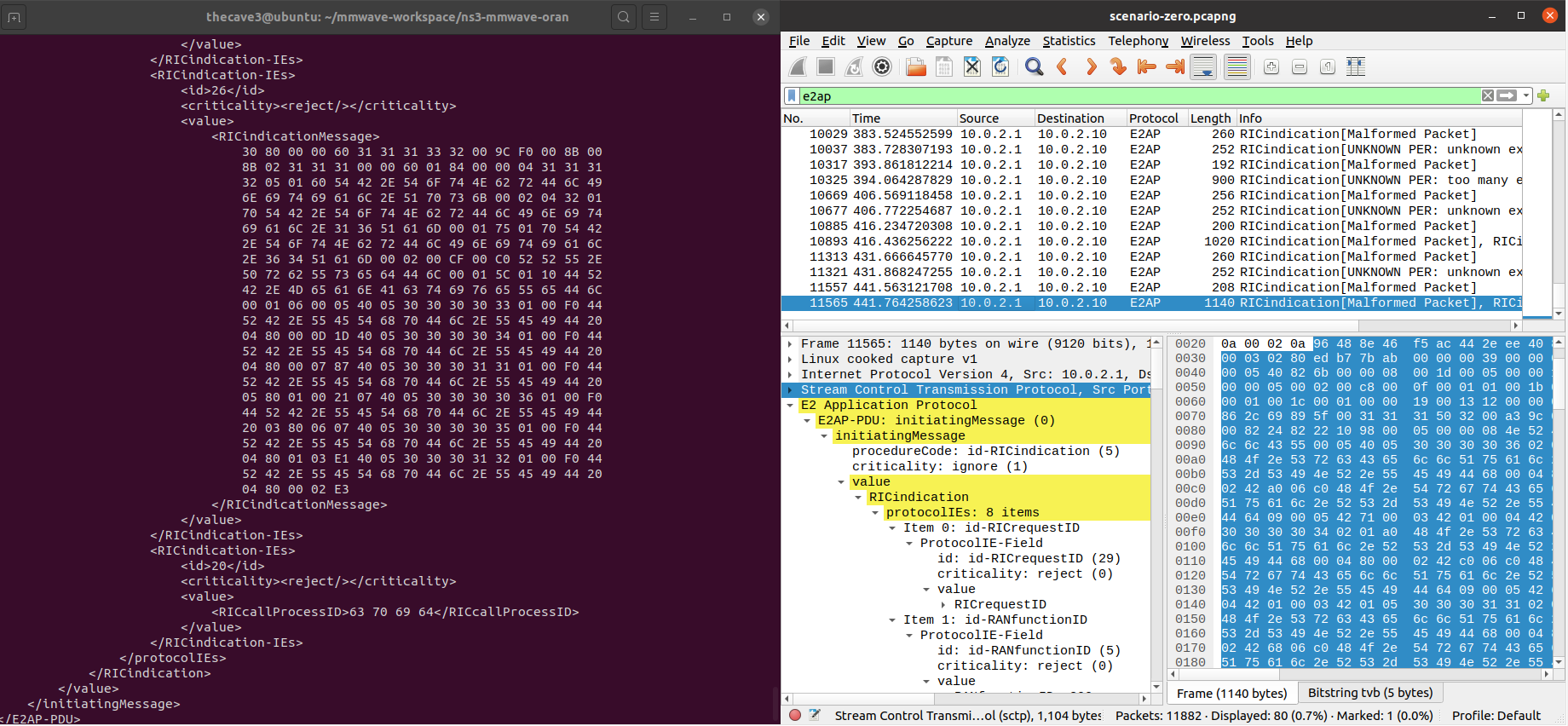}
    \caption{Dissected Packets from the Sample Scenario}
    \label{fig:wireshark}
\end{figure}

Moreover, it is possible to observe and capture the traffic on the host's interfaces with common packet dissectors such as \textit{Wireshark} and \textit{tcpdump}.
This displays all the interactions between the near-RT \gls{ric} and the simulated nodes, which can be identified on Wireshark by their IP address and port pair and by their \gls{gnb}-ID in the \gls{e2ap} protocol.
An example of this is shown in Figure~\ref{fig:wireshark}, whose left part shows the Indication Message sent by ns-3 and whose right part reports the same packet captured and dissected in \texttt{Wireshark}.
By further exploiting the packet dissector, at the end of the simulation, we can compute the statistics of the overall E2 traffic between the hosts to provide some metrics that can be useful to better understand the bandwidth usage of the \gls{e2ap} exchanges. Table~\ref{tab:recapzero} summarizes the \gls{e2ap} packets collected during the execution of Scenario Zero and gives some general statistics about its consumption from a network usage point of view.
It is important to specify that the time span is the real time duration of the simulation since we are analyzing real exchanged packets with \texttt{Wireshark} but instantiated by \gls{ns3}. From the number of bytes exchanged we notice that the traffic, which includes bytes exchanged, average data rate expressed in Bytes per second (Bps) and bit per second (bps), generated by the \gls{enb} is small if compared to the \gls{gnb}. This is mostly because the 5G \gls{bs} also provides the E2 \gls{du} Indication reports.
 
\textbf{Example use case.} 
We have used ns-O-RAN and the \gls{e2sm} implemented to study the O-RAN \gls{ts} use case~\cite{oran-wg3-use-cases} in our work~\cite{lacava2022programmable}, where we leverage ns-O-RAN and the Mavenir near-RT RIC to deploy an \gls{ai}-driven solution based on \gls{dqn} for the control of handover in an \gls{nsa} \gls{5g} simulated \gls{ran} with seven NR \glspl{gnb} and one \gls{lte} base station. This is an example of how ns-O-RAN can be used to study the O-RAN architecture and to enable the development of practical and effective \gls{ai} solutions on open cellular networks.
In the same work, we prove how ns-O-RAN can scale in complex scenarios with eight cellular \gls{bs} and more than one hundred \glspl{ue} connected and generating data which is subsequently reported to the near-RT \gls{ric}.

\section{Conclusions and Future Works}
\label{sec:conclusions}

In this paper, we presented ns-O-RAN, a framework for the integration of 5G simulations in ns-3 and near-real-time \glspl{ric}. It enables the development of xApps on real-world platforms and the testing of their capabilities on a simulated \gls{ran}, as well as extensive data collection campaigns to generate datasets for the training of AI/ML components in the xApps themselves. We discussed the main components of ns-O-RAN, and reviewed the instructions on how it can be configured and instantiated. We also introduced sample results on E2 traffic and described different methods to interact with message exchanges in the E2 termination.

As part of our future work, we will continue to improve and extend ns-O-RAN, which is a project part of the O-RAN Alliance \gls{osc} group. 
Currently, ns-O-RAN only ingests handover control messages.
We plan to extend the E2SM RC implementation support to more control actions related, for example, to \gls{pdcp} split control and slicing. 
In addition, we will upgrade the ns-3 module to the latest release, which supports the new Cmake-based build system, and merge the extension of the 4G and 5G \texttt{NetDevices} in the upstream ns-3 mmWave module. 
%
%
\balance
Finally, we will keep updating the \gls{asn1} definitions of the \gls{e2ap} and \gls{e2sm} to the latest O-RAN Alliance specifications, and consider introducing a research-oriented E2SM implemented through JSON schemas or protobufs to simplify the development of new use cases without the need to extend ASN.1 definitions.


\begin{acks}
This work was partially supported by Mavenir, by Sapienza, University of Rome under Grant "Progetti per Avvio alla Ricerca - Tipo 1", year 2022 (AR1221816B3DC365), and by the U.S. National Science Foundation under Grants CNS-1923789 and CNS-2112471.
At the time the research was conducted, Tommaso Zugno was affiliated with Northeastern University and University of Padova.
\end{acks}

\bibliographystyle{ACM-Reference-Format.bst}
\bibliography{bibl.bib}

\end{document}

%% file: acronyms.tex
\newacronym{3gpp}{3GPP}{3rd Generation Partnership Project}
\newacronym{4g}{4G}{4th generation}
\newacronym{5g}{5G}{5th generation}
\newacronym{6g}{6G}{6th generation}
\newacronym{5gc}{5GC}{5G Core}
\newacronym{adc}{ADC}{Analog to Digital Converter}
\newacronym{aerpaw}{AERPAW}{Aerial Experimentation and Research Platform for Advanced Wireless}
\newacronym{ai}{AI}{Artificial Intelligence}
\newacronym{aimd}{AIMD}{Additive Increase Multiplicative Decrease}
\newacronym{am}{AM}{Acknowledged Mode}
\newacronym{amc}{AMC}{Adaptive Modulation and Coding}
\newacronym{amf}{AMF}{Access and Mobility Management Function}
\newacronym{aops}{AOPS}{Adaptive Order Prediction Scheduling}
\newacronym{api}{API}{Application Programming Interface}
\newacronym{apn}{APN}{Access Point Name}
\newacronym{ap}{AP}{Application Protocol}
\newacronym{aqm}{AQM}{Active Queue Management}
\newacronym{asn1}{ASN.1}{Abstract Syntax Notation One}
\newacronym{ausf}{AUSF}{Authentication Server Function}
\newacronym{avc}{AVC}{Advanced Video Coding}
\newacronym{awgn}{AGWN}{Additive White Gaussian Noise}
\newacronym{balia}{BALIA}{Balanced Link Adaptation Algorithm}
\newacronym{bbu}{BBU}{Base Band Unit}
\newacronym{bdp}{BDP}{Bandwidth-Delay Product}
\newacronym{ber}{BER}{Bit Error Rate}
\newacronym{bf}{BF}{Beamforming}
\newacronym{bler}{BLER}{Block Error Rate}
\newacronym{brr}{BRR}{Bayesian Ridge Regressor}
\newacronym{bs}{BS}{Base Station}
\newacronym{bsr}{BSR}{Buffer Status Report}
\newacronym{bss}{BSS}{Business Support System}
\newacronym{ca}{CA}{Carrier Aggregation}
\newacronym{caas}{CaaS}{Connectivity-as-a-Service}
\newacronym{cb}{CB}{Code Block}
\newacronym{cc}{CC}{Congestion Control}
\newacronym{compc}{CC}{Component Carrier}
\newacronym{ccid}{CCID}{Congestion Control ID}
\newacronym{cco}{CC}{Carrier Component}
\newacronym{cdd}{CDD}{Cyclic Delay Diversity}
\newacronym{cdf}{CDF}{Cumulative Distribution Function}
\newacronym{cdn}{CDN}{Content Distribution Network}
\newacronym{cn}{CN}{Core Network}
\newacronym{cnn}{CNN}{Convolutional Neural Network}
\newacronym{codel}{CoDel}{Controlled Delay Management}
\newacronym{comac}{COMAC}{Converged Multi-Access and Core}
\newacronym{cord}{CORD}{Central Office Re-architected as a Datacenter}
\newacronym{cornet}{CORNET}{COgnitive Radio NETwork}
\newacronym{cosmos}{COSMOS}{Cloud Enhanced Open Software Defined Mobile Wireless Testbed for City-Scale Deployment}
\newacronym{cots}{COTS}{Commercial Off-the-Shelf}
\newacronym{cp}{CP}{Control Plane}
\newacronym{cpu}{CPU}{Central Processing Unit}
\newacronym{cqi}{CQI}{Channel Quality Information}
\newacronym{cr}{CR}{Cognitive Radio}
\newacronym{cql}{CQL}{Conservative Q-Learning}
\newacronym{cran}{CRAN}{Cloud \gls{ran}}
\newacronym{crs}{CRS}{Cell Reference Signal}
\newacronym{csi}{CSI}{Channel State Information}
\newacronym{csirs}{CSI-RS}{Channel State Information - Reference Signal}
\newacronym{cu}{CU}{Centralized Unit}
\newacronym{cucp}{CU-CP}{Centralized Unit - Control Plane}
\newacronym{cuup}{CU-UP}{Centralized Unit - User Plane}
\newacronym{d2tcp}{D$^2$TCP}{Deadline-aware Data center TCP}
\newacronym{d3}{D$^3$}{Deadline-Driven Delivery}
\newacronym{dac}{DAC}{Digital to Analog Converter}
\newacronym{dag}{DAG}{Directed Acyclic Graph}
\newacronym{das}{DAS}{Distributed Antenna System}
\newacronym{dash}{DASH}{Dynamic Adaptive Streaming over HTTP}
\newacronym{dc}{DC}{Dual Connectivity}
\newacronym{dccp}{DCCP}{Datagram Congestion Control Protocol}
\newacronym{dce}{DCE}{Direct Code Execution}
\newacronym{dci}{DCI}{Downlink Control Information}
\newacronym{dctcp}{DCTCP}{Data Center TCP}
\newacronym{dl}{DL}{Downlink}
\newacronym{dmr}{DMR}{Deadline Miss Ratio}
\newacronym{dmrs}{DMRS}{DeModulation Reference Signal}
\newacronym{drlcc}{DRL-CC}{Deep Reinforcement Learning Congestion Control}
\newacronym{drb}{DRB}{Data Radio Bearer}
\newacronym{drs}{DRS}{Discovery Reference Signal}
\newacronym{dnn}{DNN}{Deep Neural Network}
\newacronym{dqn}{DQN}{Deep Q-Network}
\newacronym{du}{DU}{Distributed Unit}
\newacronym{e2e}{E2E}{end-to-end}
\newacronym{e2ap}{E2AP}{E2 Application Protocol}
\newacronym{e2sm}{E2SM}{E2 Service Model}
\newacronym{ecaas}{ECaaS}{Edge-Cloud-as-a-Service}
\newacronym{ecn}{ECN}{Explicit Congestion Notification}
\newacronym{edc}{EDC}{Edge Data Center}
\newacronym{edf}{EDF}{Earliest Deadline First}
\newacronym{embb}{eMBB}{Enhanced Mobile Broadband}
\newacronym{empower}{EMPOWER}{EMpowering transatlantic PlatfOrms for advanced WirEless Research}
\newacronym{enb}{eNB}{evolved Node Base}
\newacronym{endc}{EN-DC}{E-UTRAN-NR Dual Connectivity}
\newacronym{epc}{EPC}{Evolved Packet Core}
\newacronym{eps}{EPS}{Evolved Packet System}
\newacronym{es}{ES}{Edge Server}
\newacronym{etl}{ETL}{Extract, Transform and Load}
\newacronym{etsi}{ETSI}{European Telecommunications Standards Institute}
\newacronym[firstplural=Estimated Times of Arrival (ETAs)]{eta}{ETA}{Estimated Time of Arrival}
\newacronym{eutran}{E-UTRAN}{Evolved Universal Terrestrial Access Network}
\newacronym{faas}{FaaS}{Function-as-a-Service}
\newacronym{fapi}{FAPI}{Functional Application Platform Interface}
\newacronym{fcaps}{FCAPS}{Fault, Configuration, Accounting, Performance and Security}
\newacronym{fdd}{FDD}{Frequency Division Duplexing}
\newacronym{fdm}{FDM}{Frequency Division Multiplexing}
\newacronym{fdma}{FDMA}{Frequency Division Multiple Access}
\newacronym{fed4fire}{FED4FIRE+}{Federation 4 Future Internet Research and Experimentation Plus}
\newacronym{fir}{FIR}{Finite Impulse Response}
\newacronym{fit}{FIT}{Future \acrlong{iot}}
\newacronym{fpga}{FPGA}{Field Programmable Gate Array}
\newacronym{fr1}{FR1}{Frequency Range 1}
\newacronym{fr2}{FR2}{Frequency Range 2}
\newacronym{fs}{FS}{Fast Switching}
\newacronym{fscc}{FSCC}{Flow Sharing Congestion Control}
\newacronym{ftp}{FTP}{File Transfer Protocol}
\newacronym{fw}{FW}{Flow Window}
\newacronym{gbr}{GBR}{Guaranteed Bit Rate}
\newacronym{ge}{GE}{Gaussian Elimination}
\newacronym{gnb}{gNB}{Next Generation Node Base}
\newacronym{gop}{GOP}{Group of Pictures}
\newacronym{gpr}{GPR}{Gaussian Process Regressor}
\newacronym{gpu}{GPU}{Graphics Processing Unit}
\newacronym{gtp}{GTP}{GPRS Tunneling Protocol}
\newacronym{gtpc}{GTP-C}{GPRS Tunnelling Protocol Control Plane}
\newacronym{gtpu}{GTP-U}{GPRS Tunnelling Protocol User Plane}
\newacronym{gtpv2c}{GTPv2-C}{\gls{gtp} v2 - Control}
\newacronym{gw}{GW}{Gateway}
\newacronym{harq}{HARQ}{Hybrid Automatic Repeat reQuest}
\newacronym{hetnet}{HetNet}{Heterogeneous Network}
\newacronym{hh}{HH}{Hard Handover}
\newacronym{ho}{HO}{Handover}
\newacronym{hol}{HOL}{Head-of-Line}
\newacronym{hqf}{HQF}{Highest-quality-first}
\newacronym{hss}{HSS}{Home Subscription Server}
\newacronym{http}{HTTP}{HyperText Transfer Protocol}
\newacronym{ia}{IA}{Initial Access}
\newacronym{iab}{IAB}{Integrated Access and Backhaul}
\newacronym{ic}{IC}{Incident Command}
\newacronym{ietf}{IETF}{Internet Engineering Task Force}
\newacronym{imsi}{IMSI}{International Mobile Subscriber Identity}
\newacronym{imt}{IMT}{International Mobile Telecommunication}
\newacronym{iot}{IoT}{Internet of Things}
\newacronym{ip}{IP}{Internet Protocol}
\newacronym{itu}{ITU}{International Telecommunication Union}
\newacronym{kpi}{KPI}{Key Performance Indicator}
\newacronym{kpm}{KPM}{Key Performance Measurement}
\newacronym{kvm}{KVM}{Kernel-based Virtual Machine}
\newacronym{los}{LOS}{Line-of-Sight}
\newacronym{ldc}{LDC}{Local Data Center}
\newacronym{lsm}{LSM}{Link-to-System Mapping}
\newacronym{lstm}{LSTM}{Long Short Term Memory}
\newacronym{lte}{LTE}{Long Term Evolution}
\newacronym{lxc}{LXC}{Linux Container}
\newacronym{m2m}{M2M}{Machine to Machine}
\newacronym{mac}{MAC}{Medium Access Control}
\newacronym{manet}{MANET}{Mobile Ad Hoc Network}
\newacronym{mano}{MANO}{Management and Orchestration}
\newacronym{mbr}{MBR}{Maximum Bit Rate}
\newacronym{mc}{MC}{Multi-Connectivity}
\newacronym{mcc}{MCC}{Mobile Cloud Computing}
\newacronym{mchem}{MCHEM}{Massive Channel Emulator}
\newacronym{mcs}{MCS}{Modulation and Coding Scheme}
\newacronym{mdp}{MDP}{Markov Decision Process}
\newacronym{mec}{MEC}{Multi-access Edge Computing}
\newacronym{mec2}{MEC}{Mobile Edge Cloud}
\newacronym{mfc}{MFC}{Mobile Fog Computing}
\newacronym{mgen}{MGEN}{Multi-Generator}
\newacronym{mi}{MI}{Mutual Information}
\newacronym{mib}{MIB}{Master Information Block}
\newacronym{miesm}{MIESM}{Mutual Information Based Effective SINR}
\newacronym{mimo}{MIMO}{Multiple Input, Multiple Output}
\newacronym{ml}{ML}{Machine Learning}
\newacronym{mlr}{MLR}{Maximum-local-rate}
\newacronym[plural=\gls{mme}s,firstplural=Mobility Management Entities (MMEs)]{mme}{MME}{Mobility Management Entity}
\newacronym{mmtc}{mMTC}{Massive Machine-Type Communications}
\newacronym{mmwave}{mmWave}{millimeter wave}
\newacronym{mpdccp}{MP-DCCP}{Multipath Datagram Congestion Control Protocol}
\newacronym{mptcp}{MPTCP}{Multipath TCP}
\newacronym{mr}{MR}{Maximum Rate}
\newacronym{mrdc}{MR-DC}{Multi \gls{rat} \gls{dc}}
\newacronym{mse}{MSE}{Mean Square Error}
\newacronym{mss}{MSS}{Maximum Segment Size}
\newacronym{mt}{MT}{Mobile Termination}
\newacronym{mtd}{MTD}{Machine-Type Device}
\newacronym{mtu}{MTU}{Maximum Transmission Unit}
\newacronym{mumimo}{MU-MIMO}{Multi-user \gls{mimo}}
\newacronym{mvno}{MVNO}{Mobile Virtual Network Operator}
\newacronym{nalu}{NALU}{Network Abstraction Layer Unit}
\newacronym{nas}{NAS}{Non-Access Stratum}
\newacronym{ngran}{NG-RAN}{Next Generation - \gls{ran}}
\newacronym{ns3}{ns-3}{Network Simulator 3}
\newacronym{nbiot}{NB-IoT}{Narrow Band IoT}
\newacronym{nf}{NF}{Network Function}
\newacronym{nfv}{NFV}{Network Function Virtualization}
\newacronym{nfvi}{NFVI}{Network Function Virtualization Infrastructure}
\newacronym{nic}{NIC}{Network Interface Card}
\newacronym{nlos}{NLOS}{Non-Line-of-Sight}
\newacronym{now}{NOW}{Non Overlapping Window}
\newacronym{nsm}{NSM}{Network Service Mesh}
\newacronym{nrf}{NRF}{Network Repository Function}
\newacronym{nsa}{NSA}{Non Stand Alone}
\newacronym{nse}{NSE}{Network Slicing Engine}
\newacronym{nssf}{NSSF}{Network Slice Selection Function}
\newacronym{o2i}{O2I}{Outdoor to Indoor}
\newacronym{oai}{OAI}{OpenAirInterface}
\newacronym{oaicn}{OAI-CN}{\gls{oai} \acrlong{cn}}
\newacronym{oairan}{OAI-RAN}{\acrlong{oai} \acrlong{ran}}
\newacronym{oam}{OAM}{Operations, Administration and Maintenance}
\newacronym{ofdm}{OFDM}{Orthogonal Frequency Division Multiplexing}
\newacronym{olia}{OLIA}{Opportunistic Linked Increase Algorithm}
\newacronym{omec}{OMEC}{Open Mobile Evolved Core}
\newacronym{onap}{ONAP}{Open Network Automation Platform}
\newacronym{onf}{ONF}{Open Networking Foundation}
\newacronym{onos}{ONOS}{Open Networking Operating System}
\newacronym{oom}{OOM}{\gls{onap} Operations Manager}
\newacronym{opnfv}{OPNFV}{Open Platform for \gls{nfv}}
\newacronym{orbit}{ORBIT}{Open-Access Research Testbed for Next-Generation Wireless Networks}
\newacronym{os}{OS}{Operating System}
\newacronym{oss}{OSS}{Operations Support System}
\newacronym{pa}{PA}{Position-aware}
\newacronym{pase}{PASE}{Prioritization, Arbitration, and Self-adjusting Endpoints}
\newacronym{pawr}{PAWR}{Platforms for Advanced Wireless Research}
\newacronym{pbch}{PBCH}{Physical Broadcast Channel}
\newacronym{pcell}{PCell}{Primary Cell}
\newacronym{pcef}{PCEF}{Policy and Charging Enforcement Function}
\newacronym{pcfich}{PCFICH}{Physical Control Format Indicator Channel}
\newacronym{pcrf}{PCRF}{Policy and Charging Rules Function}
\newacronym{pdcch}{PDCCH}{Physical Downlink Control Channel}
\newacronym{pdcp}{PDCP}{Packet Data Convergence Protocol}
\newacronym{pdcpc}{PDCP-C}{Packet Data Convergence Protocol - Control Plane}
\newacronym{pdcpu}{PDCP-U}{Packet Data Convergence Protocol - User Plane}
\newacronym{pdsch}{PDSCH}{Physical Downlink Shared Channel}
\newacronym{pdu}{PDU}{Packet Data Unit}
\newacronym{pf}{PF}{Proportional Fair}
\newacronym{pgw}{PGW}{Packet Gateway}
\newacronym{phich}{PHICH}{Physical Hybrid ARQ Indicator Channel}
\newacronym{phy}{PHY}{Physical}
\newacronym{phyu}{PHY-U}{Upper Physical}
\newacronym{phyl}{PHY-L}{Lower Physical}
\newacronym{pmch}{PMCH}{Physical Multicast Channel}
\newacronym{pmi}{PMI}{Precoding Matrix Indicators}
\newacronym{powder}{POWDER}{Platform for Open Wireless Data-driven Experimental Research}
\newacronym{ppo}{PPO}{Proximal Policy Optimization}
\newacronym{ppp}{PPP}{Poisson Point Process}
\newacronym{prach}{PRACH}{Physical Random Access Channel}
\newacronym{prb}{PRB}{Physical Resource Block}
\newacronym{pscell}{PSCell}{Primary cell of the Secondary Node}
\newacronym{psnr}{PSNR}{Peak Signal to Noise Ratio}
\newacronym{pss}{PSS}{Primary Synchronization Signal}
\newacronym{pucch}{PUCCH}{Physical Uplink Control Channel}
\newacronym{pusch}{PUSCH}{Physical Uplink Shared Channel}
\newacronym{qam}{QAM}{Quadrature Amplitude Modulation}
\newacronym{qci}{QCI}{\gls{qos} Class Identifier}
\newacronym{qoe}{QoE}{Quality of Experience}
\newacronym{qos}{QoS}{Quality of Service}
\newacronym{quic}{QUIC}{Quick UDP Internet Connections}
\newacronym{rach}{RACH}{Random Access Channel}
\newacronym{ran}{RAN}{Radio Access Network}
\newacronym[firstplural=Radio Access Technologies (RATs)]{rat}{RAT}{Radio Access Technology}
\newacronym{rcn}{RCN}{Research Coordination Network}
\newacronym{rec}{REC}{Radio Edge Cloud}
\newacronym{red}{RED}{Random Early Detection}
\newacronym{rem}{REM}{Random Ensemble Mixture}
\newacronym{renew}{RENEW}{Reconfigurable Eco-system for Next-generation End-to-end Wireless}
\newacronym{rf}{RF}{Radio Frequency}
\newacronym{rfc}{RFC}{Request for Comments}
\newacronym{rfr}{RFR}{Random Forest Regressor}
\newacronym{ric}{RIC}{RAN Intelligent Controller}
\newacronym{rlc}{RLC}{Radio Link Control}
\newacronym{rl}{RL}{Reinforcement Learning}
\newacronym{rlf}{RLF}{Radio Link Failure}
\newacronym{rlnc}{RLNC}{Random Linear Network Coding}
\newacronym{rmr}{RMR}{RIC Message Routing}
\newacronym{rmse}{RMSE}{Root Mean Squared Error}
\newacronym{rnis}{RNIS}{Radio Network Information Service}
\newacronym{rnib}{RNIB}{Radio Network Information Base}
\newacronym{rr}{RR}{Round Robin}
\newacronym{rrc}{RRC}{Radio Resource Control}
\newacronym{rrm}{RRM}{Radio Resource Management}
\newacronym{rru}{RRU}{Remote Radio Unit}
\newacronym{rs}{RS}{Remote Server}
\newacronym{rsrp}{RSRP}{Reference Signal Received Power}
\newacronym{rsrq}{RSRQ}{Reference Signal Received Quality}
\newacronym{rss}{RSS}{Received Signal Strength}
\newacronym{rssi}{RSSI}{Received Signal Strength Indicator}
\newacronym{rtt}{RTT}{Round Trip Time}
\newacronym{ru}{RU}{Radio Unit}
\newacronym{rw}{RW}{Receive Window}
\newacronym{rx}{RX}{Receiver}
\newacronym{s1ap}{S1AP}{S1 Application Protocol}
\newacronym{sa}{SA}{standalone}
\newacronym{sack}{SACK}{Selective Acknowledgment}
\newacronym{sap}{SAP}{Service Access Point}
\newacronym{sc2}{SC2}{Spectrum Collaboration Challenge}
\newacronym{scef}{SCEF}{Service Capability Exposure Function}
\newacronym{sch}{SCH}{Secondary Cell Handover}
\newacronym{scoot}{SCOOT}{Split Cycle Offset Optimization Technique}
\newacronym{sctp}{SCTP}{Stream Control Transmission Protocol}
\newacronym{sdap}{SDAP}{Service Data Adaptation Protocol}
\newacronym{sdk}{SDK}{Software Development Kit}
\newacronym{sdm}{SDM}{Space Division Multiplexing}
\newacronym{sdma}{SDMA}{Spatial Division Multiple Access}
\newacronym{sdn}{SDN}{Software-defined Networking}
\newacronym{sdr}{SDR}{Software-defined Radio}
\newacronym{seba}{SEBA}{SDN-Enabled Broadband Access}
\newacronym{sgsn}{SGSN}{Serving GPRS Support Node}
\newacronym{sgw}{SGW}{Serving Gateway}
\newacronym{si}{SI}{Study Item}
\newacronym{sib}{SIB}{Secondary Information Block}
\newacronym{sinr}{SINR}{Signal to Interference plus Noise Ratio}
\newacronym{sip}{SIP}{Session Initiation Protocol}
\newacronym{siso}{SISO}{Single Input, Single Output}
\newacronym{sla}{SLA}{Service Level Agreement}
\newacronym{sm}{SM}{Service Model}
\newacronym{smf}{SMF}{Session Management Function}
\newacronym{smo}{SMO}{Service Management and Orchestration}
\newacronym{sms}{SMS}{Short Message Service}
\newacronym{smsgmsc}{SMS-GMSC}{\gls{sms}-Gateway}
\newacronym{snr}{SNR}{Signal-to-Noise-Ratio}
\newacronym{son}{SON}{Self-Organizing Network}
\newacronym{sptcp}{SPTCP}{Single Path TCP}
\newacronym{srb}{SRB}{Service Radio Bearer}
\newacronym{srn}{SRN}{Standard Radio Node}
\newacronym{srs}{SRS}{Sounding Reference Signal}
\newacronym{ss}{SS}{Synchronization Signal}
\newacronym{sss}{SSS}{Secondary Synchronization Signal}
\newacronym{st}{ST}{Spanning Tree}
\newacronym{svc}{SVC}{Scalable Video Coding}
\newacronym{tb}{TB}{Transport Block}
\newacronym{tcp}{TCP}{Transmission Control Protocol}
\newacronym{tdd}{TDD}{Time Division Duplexing}
\newacronym{tdm}{TDM}{Time Division Multiplexing}
\newacronym{tdma}{TDMA}{Time Division Multiple Access}
\newacronym{tfl}{TfL}{Transport for London}
\newacronym{tfrc}{TFRC}{TCP-Friendly Rate Control}
\newacronym{tft}{TFT}{Traffic Flow Template}
\newacronym{tgen}{TGEN}{Traffic Generator}
\newacronym{tip}{TIP}{Telecom Infra Project}
\newacronym{tm}{TM}{Transparent Mode}
\newacronym{to}{TO}{Telco Operator}
\newacronym{tr}{TR}{Technical Report}
\newacronym{trp}{TRP}{Transmitter Receiver Pair}
\newacronym{ts}{TS}{Traffic Steering}
\newacronym{tti}{TTI}{Transmission Time Interval}
\newacronym{ttt}{TTT}{Time-to-Trigger}
\newacronym{tx}{TX}{Transmitter}
\newacronym{uas}{UAS}{Unmanned Aerial System}
\newacronym{uav}{UAV}{Unmanned Aerial Vehicle}
\newacronym{udm}{UDM}{Unified Data Management}
\newacronym{udp}{UDP}{User Datagram Protocol}
\newacronym{udr}{UDR}{Unified Data Repository}
\newacronym{ue}{UE}{User Equipment}
\newacronym{uhd}{UHD}{\gls{usrp} Hardware Driver}
\newacronym{ul}{UL}{Uplink}
\newacronym{um}{UM}{Unacknowledged Mode}
\newacronym{uml}{UML}{Unified Modeling Language}
\newacronym{upa}{UPA}{Uniform Planar Array}
\newacronym{upf}{UPF}{User Plane Function}
\newacronym{urllc}{URLLC}{Ultra Reliable and Low Latency Communications}
\newacronym{usa}{U.S.}{United States}
\newacronym{usim}{USIM}{Universal Subscriber Identity Module}
\newacronym{usrp}{USRP}{Universal Software Radio Peripheral}
\newacronym{utc}{UTC}{Urban Traffic Control}
\newacronym{vim}{VIM}{Virtualization Infrastructure Manager}
\newacronym{vm}{VM}{Virtual Machine}
\newacronym{vnf}{vNF}{Virtualized Network Function}
\newacronym{volte}{VoLTE}{Voice over \gls{lte}}
\newacronym{voltha}{VOLTHA}{Virtual OLT HArdware Abstraction}
\newacronym{vr}{VR}{Virtual Reality}
\newacronym{vran}{vRAN}{Virtualized \gls{ran}}
\newacronym{vss}{VSS}{Video Streaming Server}
\newacronym{v2x}{V2X}{Vehicle-to-everything}
\newacronym{wbf}{WBF}{Wired Bias Function}
\newacronym{wf}{WF}{Waterfilling}
\newacronym{wlan}{WLAN}{Wireless Local Area Network}
\newacronym{osm}{OSM}{Open Source \gls{nfv} Management and Orchestration}
\newacronym{pnf}{PNF}{Physical Network Function}
\newacronym{drl}{DRL}{Deep Reinforcement Learning}
\newacronym{mtc}{MTC}{Machine-type Communications}
\newacronym{osc}{OSC}{O-RAN Software Community}
\newacronym{rc}{RC}{RAN Control}
\newacronym{ar}{AR}{Augmented Reality}
\newacronym{daps}{DAPS}{Dual Active Protocol Stack}
\newacronym{nib}{NIB}{Network Information Base}

%% file: myTikz.tex
\usepackage{tikz}
\usepackage{pgfplots}
\pgfplotsset{compat=newest} 
\pgfplotsset{plot coordinates/math parser=false} 
\newlength\fheight
\newlength\fwidth
\usetikzlibrary{plotmarks,patterns,decorations.pathreplacing,backgrounds,calc,arrows,arrows.meta,spy,matrix}
\usepgfplotslibrary{patchplots,groupplots}
\usepackage{tikzscale}

\tikzstyle{startstop} = [rectangle, rounded corners, minimum width=2cm, minimum height=0.5cm,text centered, draw=black]
\tikzstyle{io} = [trapezium, trapezium left angle=70, trapezium right angle=110, minimum width=3cm, minimum height=1cm, text centered, draw=black]
\tikzstyle{process} = [rectangle, minimum width=2cm, minimum height=0.5cm, text centered, draw=black, align=center]
\tikzstyle{decision} = [ellipse, minimum width=2cm, minimum height=1cm, text centered, draw=black]
\tikzstyle{arrow} = [thick,<->,>=stealth]
\tikzstyle{line} = [thick,>=stealth]
\tikzstyle{darrow} = [thick,<->,>=stealth,dashed]
\tikzstyle{sarrow} = [thick,->,>=stealth]
\tikzstyle{larrow} = [line width=0.1mm,dashdotted,<->,>=stealth]

\definecolor{SchoolColor}{RGB}{0.71, 0, 0.106}
\definecolor{chaptergrey}{rgb}{0.61, 0, 0.09} 
\definecolor{midgrey}{rgb}{0.4, 0.4, 0.4}
\definecolor{chaptergreen}{rgb}{0.09, 0.612, 0}
\definecolor{chapterpurple}{rgb}{0.522, 0, 0.612}
\definecolor{chapterlightgreen}{rgb}{0, 0.612, 0.522}

\makeatletter
\def\grd@save@target#1{%
  \def\grd@target{#1}}
\def\grd@save@start#1{%
  \def\grd@start{#1}}
\tikzset{
  grid with coordinates/.style={
    to path={%
      \pgfextra{%
        \edef\grd@@target{(\tikztotarget)}%
        \tikz@scan@one@point\grd@save@target\grd@@target\relax
        \edef\grd@@start{(\tikztostart)}%
        \tikz@scan@one@point\grd@save@start\grd@@start\relax
        \draw[minor help lines] (\tikztostart) grid (\tikztotarget);
        \draw[major help lines] (\tikztostart) grid (\tikztotarget);
        \grd@start
        \pgfmathsetmacro{\grd@xa}{\the\pgf@x/1cm}
        \pgfmathsetmacro{\grd@ya}{\the\pgf@y/1cm}
        \grd@target
        \pgfmathsetmacro{\grd@xb}{\the\pgf@x/1cm}
        \pgfmathsetmacro{\grd@yb}{\the\pgf@y/1cm}
        \pgfmathsetmacro{\grd@xc}{\grd@xa + \pgfkeysvalueof{/tikz/grid with coordinates/major step x}}
        \pgfmathsetmacro{\grd@yc}{\grd@ya + \pgfkeysvalueof{/tikz/grid with coordinates/major step y}}
        \foreach \x in {\grd@xa,\grd@xc,...,\grd@xb}
        \node[anchor=north] at (\x,\grd@ya) {\pgfmathprintnumber{\x}};
        \foreach \y in {\grd@ya,\grd@yc,...,\grd@yb}
        \node[anchor=east] at (\grd@xa,\y) {\pgfmathprintnumber{\y}};
      }
    }
  },
  minor help lines/.style={
    help lines,
    gray,
    line cap =round,
    xstep=\pgfkeysvalueof{/tikz/grid with coordinates/minor step x},
    ystep=\pgfkeysvalueof{/tikz/grid with coordinates/minor step y}
  },
  major help lines/.style={
    help lines,
    line cap =round,
    line width=\pgfkeysvalueof{/tikz/grid with coordinates/major line width},
    xstep=\pgfkeysvalueof{/tikz/grid with coordinates/major step x},
    ystep=\pgfkeysvalueof{/tikz/grid with coordinates/major step y}
  },
  grid with coordinates/.cd,
  minor step x/.initial=.5,
  minor step y/.initial=.2,
  major step x/.initial=1,
  major step y/.initial=1,
  major line width/.initial=1pt,
}
\makeatother